\documentclass[aps,12pt,final,notitlepage,oneside,onecolumn,nobibnotes,nofootinbib,superscriptaddress,noshowpacs]{revtex4}

 \usepackage{graphicx}

\usepackage{amssymb}

\usepackage[english,francais]{babel}

\newtheorem{e-proposition}[theorem]{Proposition}

\newtheorem{e-definition}[theorem]{Definition\rm}

\def\og{\leavevmode\raise.3ex\hbox{$\scriptscriptstyle\langle\!\langle$~}}
\def\fg{\leavevmode\raise.3ex\hbox{~$\!\scriptscriptstyle\,\rangle\!\rangle$}}

\begin{document}




\title{Herzfeld versus Mott transition in metal-ammonia solutions.}
\author{G.N. Chuev$^{(1,2)}$, P. Qu\'emerais$^{(3)}$}

\affiliation{$^{(1)}$ Institute of Theoretical and Experimental Biophysics, Russian Academy of Science, Pushchino, Moscow Region, 142290, Russia
and
$^{(2)}$ School of Physics, The University of Edimburgh, Mayfield Road, Edinburgh EH9 3JZ, United Kingdom \\
$^{(3)}$Institut N\'eel, CNRS, BP 166, 38042 Grenoble Cedex 9, France }







\begin{abstract}
Although most metal-insulator transitions in doped insulators are generally viewed as Mott
transitions, some systems seem to deviate from this scenario. Alkali metal-ammonia solutions are a brilliant example of that. They reveal a phase separation in the range
of metal concentrations where a metal-insulator transition occurs.
Using a mean spherical approximation for quantum polarizable fluids,
we argue that the origin of the metal-insulator transition in such a
system is likely similar to that proposed by Herzfeld a long time ago,
namely, due to fluctuations of solvated electrons. We also
show how the phase separation may appear: the Herzfeld instability of
the insulator occurs at a concentration for which the metallic
phase is also unstable. As a consequence, the Mott transition
cannot occur at low
temperatures. The proposed scenario may provide a new insight
into the metal-insulator transition in condensed-matter physics.

\vskip 0.5\baselineskip

keywords: {Solvated electrons; Mott transition; Phase separation }
\end{abstract}


\maketitle

In the first theory of metallization, Herzfeld \cite{herzfeld} has considered dipolar fluctuations of neutral atoms as the origin of a
metal-insulator transition (MIT). He emphasized that the restoring force of an electron bound to an atom collapses and the substance becomes
metallic at increased densities due to local field effects. Could the Herzfeld idea be applied to describe the MIT in real systems? Despite some
successful examples of such applications \cite{herz1,herz2,herz3,ashcroft}, serious doubts arose because most of the experiments gave evidence
in favor of the Mott scenario \cite{mott1}. The latter focused on the screening of the long-range Coulomb potential by a stable electron gas,
which prevents the formation of bound electrons. As far as this metal hallmark process runs, the system remains metallic, but when it ceases at
lower densities, a MIT occurs. It is very important to realize that the two theories are based on different effects, since Mott considered the
MIT from the metallic side of the transition, whereas Herzfeld investigated the same phenomenon from the insulating one. No comprehensive
connection between them has been provided up to now. Recent studies on a Wigner crystal formed by large polarons have however opened new
perspectives for the Herzfeld idea. It was shown \cite {quem1,rastelli} that this Wigner crystal of polarons loses its stability owing to the
dipolar interactions between polarons, yielding a polarization catastrophe, which provokes the onset of metallization. This phenomenon is a
quantum version of the classical Herzfeld scenario. Motivated by these results, we argue that a modified Herzfeld approach may provide a key to
understand the MIT in certain real systems. In particular, we focus in this paper on the metal-ammonia solutions (MAS).

\begin{figure}[tbp]
\caption{Phase diagram of Na-NH$_3$. The experimental data on the
locus of the phase separation are indicated by the square
\cite{crauss} and diamond symbols \cite{CHIEUX}, respectively.
The triangles show the change in sign of the derivative of the
conductivity coefficient $d\sigma /dT$, which is used to estimate
the locus of the MIT \cite{thompson}. The solid curve corresponds to
our calculations of the locus of the polarization catastrophe. All
dotted curves are guides for the eyes, except the dashed horizontal line at $%
T=-80^o$ C indicating the solidification temperature of ammonia.
[1(MPM) $\approx 2\cdot 10^{20}$ cm$^{-3}$].} \label{phasediagram}
\end{figure}

Although it is a century-old problem \cite{crauss}, the phase diagram
of MAS (see Fig. 1) has remained mysterious up to now. Many studies have been performed
and a large volume of experimental data have been accumulated about
this fascinating system (for review, see \cite
{thompson,edwards}). Once an alkali metal is dissolved in liquid
ammonia, it immediately dissociates to give two separated entities
with unlike charges: the solvated ions and the excess electrons. At
low metal concentration, the solution remains non metallic
(electrolytic) and has an intense blue colour independently on the
type of alkali metal. Jortner \cite{jortner1} argued that due to
short-range interactions with ammonia molecules, an excess electron
forms a cavity free of solvent in which it localizes with the help
of the polarization carried by the surrounding ammonia molecules.
This process results in a trap formation similar to that for the
polarons in solids. The radius of the cavity has been estimated to
be $r_{c}\approx 3.2$ \AA\ \cite{jortner1}. Modern
theories based on path integral simulations \cite{deng}, or on the
density functional approach \cite{chuev,Bip} provide an evaluation of the microscopic structure around
solvated electrons but they all yield the same physical picture as
that described above. At large enough metal concentration, the MAS
becomes a liquid metal with a typical bronze coloration. However, at
concentrations varying from $1$ to $10 $ mole percent of metal
(MPM), a separation between the low density blue phase and the
higher density bronze one takes place, resulting in a miscibility
gap below a critical temperature (Fig. 1). Importantly, the phase
separation occurs for Li, Na, or K, but was not observed in the case of Cs. However, for all type of alkali metal,
many experimental data reveal the presence of a MIT in the same range of
densities \cite{thompson}. This is reported in Fig. 1 for the case of Na.

Earlier models considering the Mott mechanism \cite{sienko} or
involving an association of localized electrons in clusters
\cite{mottRMP} were not able to explain the whole phase
diagram observed in MAS satisfactorily. What is the reason? Let's give an outlook
on the complexity of the problem. From an electrostatic point of
view, solvated electrons behave more or less like some solvated
anions, the counterpart of the solvated metal cations. The
Debye screening length is found to be about \AA\
at $4$ MPM (taking into account the static dielectric constant
$\epsilon_s \sim 20$ of ammonia), which makes the MAS a
strong electrolyte in this concentration range. The static Coulomb
interactions are thus essentially already screened when the MIT
occurs, and therefore cannot be its origin. The short-range
interactions between electrons are also unlikely to be responsible
for the MIT and the phase separation, because the mean distance
between electrons is still about $12$ \AA\ at the relevant
concentration $4$ MPM, which is enough to neglect any overlapping
between the wave-functions of the electrons localized in their
ground state. Finally, the occurrence of a phase separation at low
temperatures in the concentration range where the MIT occurs,
gives serious doubts about the possibility of a Mott transition. We
will come back to this point.

The origin of the MIT must be found elsewhere, and a reasonable
hypothesis is that it results from quantum momentum
fluctuation\textbf{s} of the solvated electrons. These ones are
self-trapped quantum particles, whose dipolar momentum effectively
fluctuates due to their quantum nature, with a characteristic
frequency $\omega_0(T)$. This frequency corresponds to electronic
transitions of the electrons between two states bound in their own
trap potentials. They are experimentally detected by ordinary
optical absorption measurements. The latter reveal a broad
absorption line peaked at $\omega _{0} (T) \sim 0.9$ eV at low
concentration and $T=-70^\circ$C, with a tail extending in the
visible region\emph{,} and providing the blue colour of
diluted MAS. For a given temperature, the maximum of optical
absorption shows a pronounced red shift at metal concentration above
$0.1$ MPM, which cannot be caused by short-range or static Coulomb
interactions as it has been discussed above. The frequency
$\omega_0(T)$ characterizing the
solvated electron state, is a significant phenomelogical
parameter of our theory. It is associated to the static
polarizability $\alpha_0(0) =e^{2}/m\omega _{0}^{2}\sim
10^{-22}cm^{3}$ of a solvated electron ($m$ is the electron mass).
That is a huge polarizability with respect to the one of a single
ammonia molecule $\alpha _{NH_{3}} \sim 2.8\cdot 10^{-24}cm^{3}$
\cite{ammoniapolariz} or that of a sodium ion $\alpha _{Na^{+}} \sim
2\cdot 10^{-25}cm^{3}$ \cite{sodiumpolariz}. Therefore, these
quantum fluctuations of isolated solvated electrons and their
induced dipole-dipole interactions may have a dominant role in the
MIT. Moreover, since $\omega_0 (T)$ is much higher than the inverse
relaxation time $\tau^{-1}$ of the solvent (typically in
the THz range), the dipolar interactions between solvated electrons
are only screened by the high frequency dielectric constant
$\epsilon_{\infty}$ of ammonia.

\begin{figure}[tbp]
\caption{The density of state (DOS) of the collective polarization
modes of interacting dipoles. The DOS is drawn in arbitrary units, for various metal
concentrations at $T=-35^\circ$C. The lower edge $\omega_-(T,n)$ drives
the stability of the system (see text).} \label{phasediagram}
\end{figure}

\begin{figure}[tbp]
\caption{Concentration dependencies of the dielectric
constant (a) and the locus of the maximum of optical absorption (b) in MAS. In a) the
square symbols correspond to the experimental data \cite{pk} on dielectric
constant at $T=+20^o$C,  the circle symbols to that at $T=-35^o$C
\cite{pk1}. The solid and dashed curves show our results at
$T=+20^o$C and $-35^o$C, respectively. In b) the triangle symbols indicate the
experimental data on the absorption maximum in Na-NH$_3$ at
$T=-65^o$C obtained from \cite{ccc}  and  square symbols from
\cite{ccc1}, while the solid curve shows our
results at the same temperature.}
\label{dielecwmax}
\end{figure}

The key idea of Herzfeld was to evaluate the effect of the local
field, with the help of the Clausius-Mossoti relation, as a function
of the density of a substance and the polarizability of its
constituents. Nevertheless, his calculation did not take into
account the fact that the particles interact. However, these
interactions induce collective modes of polarization, which
substantially modify the generalized susceptibility $\chi
(\omega)$ with respect to the non-interacting case. This susceptibility gives the response to an external field $\mathbf{E}_0$, i.e. ${\mathbf P}= \chi(\omega) {\mathbf E}_0$, where $\mathbf P$ is the polarization. Taking into
account both the local field and the interactions effects, we
generalize the Clausius-Mossoti relation as
\begin{equation}
\frac{\epsilon (T,\omega )/\epsilon _{NH_{3}}(T,\omega )-1}{\epsilon
(T,\omega )/\epsilon _{NH_{3}}(T,\omega )+2}=\frac{4\pi }{3} \chi( \omega) ,
\end{equation}
where $\epsilon _{NH_{3}}(T,\omega )$ is the temperature- and
frequency-dependent dielectric function of pure ammonia,
$\epsilon(T,\omega ) $ is the similar quantity of the solution. The
susceptibility may be expressed in term\textbf{s} of an effective
dynamical polarizability $\alpha(\omega)$ of a single solvated
electron, by the relation $\chi (\omega) = n \alpha(\omega)
/\epsilon_{\infty}$. Hence, the problem focuses on the calculations
of $\alpha(\omega)$. The spectral density-functional theory \cite
{Kotliar} seems to provide a rigorous basis for such calculations.
However, despite recent progresses in this direction
\cite{Kotliar1}, the complexity of the microscopic structure and the
presence of disorder in MAS prevent accurate numerical calculations
of the electronic properties.

To overcome this barrier we choose a simple semi-analytical model
suitable to calculate $\alpha (\omega )$ with a reasonable accuracy.
We consider MAS as a fluid of quantum particles localized
in cavities with diameter $\sigma =2r_{c}$, which interact through
induced dipole-dipole interactions. They can be treated as a set of
quantum Drude oscillators with isolated polarizability $\alpha
_{0}(\omega )=e^{2}/m(\omega _{0}^{2}-\omega ^{2})$. In the simplest
approximation the interactions between particles can be cut off at
small distances by the cavity size, wher\textbf{ea}s short-range
details are ignored. Similar models of quantum
polarizable fluids have been extensively studied
\cite{chandler1,shweizer,pratt,chen}, and we use the results which
were previously obtained. In particular (see Methods), the problem
is reduced to solve a quadratic equation, whose roots are complex,
i.e. $\alpha (\omega )=\alpha ^{\prime }(\omega )+i\alpha ^{\prime
\prime }(\omega )$. The imaginary part is non zero only in a finite
range of frequency $\omega _{-}(T,n)<\omega <\omega _{+}(T,n)$ that
corresponds to the dispersion of the collective polarization modes,
regarded as a distribution of eigenvalues. Their density of state
(DOS) is given by $D(\omega )\propto \omega \alpha ^{\prime \prime
}(\omega )$ \cite {chen}. Fig. 2 illustrates our calculation at
$T=-35^\circ$C. At low concentrations, the DOS is peaked at $\omega
_{0}(T)$, whereas the spectrum broadens progressively as the density
$n$ increases, indicating the drastic effect of the interactions.

Since the squared eigenfrequencies of the collective modes are to be
positive, the low edge $\omega _{-}^{2}(n,T)$ drives the stability of the
system. Generalizing the Herzfeld criterion of polarization catastrophe, we
define the critical density $n_{c1}$ of the MIT as
\begin{equation}
\omega _{-}(T,n_{c1})=0.
\end{equation}
We have reported in Fig. 1 the calculated critical densities
obtained with the use of Eq.(2). It indicates that the MIT occurs
between 2 and 5 MPM depending on the temperature, which is quite
comparable to the experimental data. The evaluation of the original
Herzfeld critical density, i.e. without taking into account the
interactions, provides the MIT located at about 14 MPM at
$T=-70^\circ$C. That shows how important the effect of the
interactions is to correctly evaluate the MIT. Another consequence of
the polarization catastrophe is that the low-frequency dielectric
constant $\epsilon (T,\omega )$ diverges at $n_{c1}$ as it is
experimentally observed in MAS \cite{pk,pk1}. Comparing the
calculated data with the experimental ones, we find a good agreement
between them (Fig. 3a). We also have calculated the real and the
imaginary parts of the dielectric constant and evaluate the optical
absorption coefficient $A(\omega )$. Again the calculated
concentration dependence of $A(\omega )$ at the locus $\omega
_{max}$ of its maximum, agrees well with the experimental data \cite
{ccc,ccc1} at concentrations below $n_{c1}$ (Fig. 3b).

\begin{figure}[tbp]
\caption{Reduced isothermal compressibility of the electron gas as a function of concentration at T=$-70^o$ C .The solid and
the dashed curves correspond to the case of Na and
Cs counterions, respectively. The arrow shows the locus of the dielectric catastrophe for the solvated electrons, indicating a miscibility gap between $n_{c1}$ and $n_{c2}(Na)$ for Na, which does not occur in the case of Cs ions (see text).} \label{instability}
\end{figure}

Coming at the issue from a different angle, let's consider
the question: could the dipolar fluctuations lead to the phase
separation experimentally observed in MAS? Above the critical concentration $n_{c1}$, the localized
electrons are not stable. Hence,
the behaviour of the system above $n_{c1}$ depends on thermodynamics
of the metallic state. But the homogeneous electron gas is
known to be unstable at sufficiently low densities due to occurrence of a
negative compressibility. Therefore, if the polarization catastrophe
occurs at a lower density than this instability, it
should provoke a phase separation. The MAS seems to be just this case. To reveal it, we have calculated
the electronic part of the compressibility for the metallic state
with the use of a modified model of stabilized jellium \cite{folias}
(see Methods) for several alkali metals. Fig. 4 shows the two curves
obtained for Na and Cs respectively. In the case of Na, the compressibility $%
\kappa_F$ diverges at a critical density $n_{c2} \approx 6$ MPM, whereas the
dielectric catastrophe of the solvated
electron state occurs at $n_{c1} \approx 5$ MPM. This is the origin of the
miscibility gap and the associated phase separation: it exists a range of
density $n \in [n_{c1},n_{c2}]$ for which both states are unstable. Another
consequence of this phenomenon is that a Mott mechanism for the MIT, which
requires a stable electron gas at the critical density, appears impossible
since the experimental MIT occurs at lower concentration than $n_{c2}$. The
second curve in Fig. 4 is for Cs. It is seen in that $n_{c2}<n_{c1}$,
contrary to the case of Na. No miscibility gap is thus expected in the case
of Cs. This result is also coherent with the experimental facts. Although our
estimations give upper and lower bounds for $n_{c1}$ and $n_{c2}$
respectively, underestimating the instability range, they reveal a correct
trend in the dependence on the size of ions, namely, a decrease of the
instability range for heavier ions, due to scattering of delocalized
electrons on ion cores. The later may decrease $n_{c2}$ enough to destroy
the miscibility gap.

In conclusion, following the Herzfeld idea and using the hard-sphere
models for quantum polarizable fluids, we have evaluated
peculiarities of MIT in MAS, namely, the anomalies of dielectric
response and concentration changes in the absorption maximum. Our estimations of the behaviour of the insulating and the metallic phases have revealed an instability
range at low temperatures. Although the predicted miscibility gap is
sufficiently smaller than the experimental one and is sensitive to
variations of the model parameters (for example, 10\% decrease in
$\epsilon _{\infty }$ enhances twice this range), our calculations
tell us that MAS system may deviate from the usual Mott scenario.
The MIT seems dominated by the old Herzfeld mechanism. Our model is
simple and does not take into account phenomena which may influence
the transition such as disorder effects on the cavity formation of
localized electrons, clusterization of electrons, influence
of ionized states of metal atoms, and so on. These effects may be
important in the vicinity of the transition, nevertheless simple
estimations yield their energy scales sufficiently lower than that
of the dipolar interactions between solvated electrons. Long-range
nature of these interactions and low density of excess electrons in
the solution support our game with the semi-analytical calculations.

We also believe that the proposed scenario may be applied to some
other systems. The reason of our optimism is the following. The
behaviour of the quantum polarizable particles with respect to the homogeneous electron gas is controlled by two
dimensionless parameters $\alpha _{0}\sigma
^{-3}$ and $n \sigma^3$. The concentrations of electrons are to be low to provide
the instability of the electron gas, and,
hence, $n \sigma ^{3}$ is to be small. But the parameter $%
\alpha _{0}\sigma ^{-3}$ is also small in the case of ordinary polarizable fluids. For such systems, the present scenario remains unlikely. However in our case of self-trapped quantum particles, $%
\alpha _{0}\sigma ^{-3}$ may be large enough to provoke a phase
separation (it is about of $0.5$ for MAS). Hence, our scenario
is general and may take place in other systems, depending on the origin of the polarizable particles. For instance, in alkali metal-alkali halide solutions, solvated electrons, phase separation and
dielectric anomalies were experimentally observed \cite{Freyland}.
Another example is the case of doped polar solids such as oxides where formation of large polarons occurs. For such materials $\alpha _{0}\sigma
^{-3}\sim \epsilon _{\infty }^{-1}-\epsilon _{s}^{-1}$ and the origin of
the unusual behaviour is essentially due to different scales of
screening at various frequencies $\epsilon _{s}>>\epsilon _{\infty
}$. We have previously revealed it  in the case of a Wigner crystal of
polarons \cite {quem1} which, from this point of view, behaves similarly like MAS. Finally,
we would like to emphasize that our scenario opens a window in the
general understanding of the MIT, facing a situation when the
metallic and the insulating states are both unstable in a finite range
of density. This stimulates the further challenging question. Can
the dipolar interactions of localized quantum particles provoke a
transition to a superconducting state? Such a possibility was already discussed in \cite{quem2}, and, perhaps,
a careful analysis of former experiments on superconductivity in frozen
MAS \cite{Ogg,Dmitrenko} could help to find the answer.

\section*{Methods}

\subsection*{Evaluations of the effective polarizability}

In the case of interacting quantum Drude oscillators the problem is
to evaluate the effective polarizability
$\alpha(\omega)$. Due to  dipolar interactions between the
oscillators, the polarizability $\alpha(\omega )$ is modified with
respect to the non-interacting case and is given by the
self-consistent equation:
\begin{equation}
\epsilon _{\infty } / \alpha \left( \omega \right) =\epsilon _{\infty
} / \alpha _{0}\left( \omega \right) - 2E(\alpha \left( \omega \right)
/\epsilon _{\infty }) ,  \label{effectivealpha}
\end{equation}
where the last term accounts the correlations between induced
dipoles. Eq.(3) has been derived in \cite{chandler1}, we only modify
it by taking into account the high-frequency screening by the solvent (the use
of $\epsilon_{\infty}$ in Eq.(3)). Formally the quantity $3\alpha
E(\alpha )/\beta $ equates with the dipolar part of the internal
energy per particle of a classical liquid of nonpolarizable
particles with permanent dipole momentum $(3\alpha /\beta )^{1/2}$
\cite{chandler1,pratt}, where $\beta $ is the inverse
temperature. Once the function $E(\alpha )$ is known, all the
physical properties of the system can be evaluated by solution of
Eq.(3). The simplest method to obtain $E(\alpha ) $ is the Pad\'{e}
approximation \cite{chandler1,shweizer}, which is an interpolation
between the case of low and large polarizability $\alpha(\omega)$.
Adapting this method to our case, we get
\begin{equation}
E(\alpha (\omega )/\epsilon _{\infty })=\frac{I_{0}(n\sigma ^{3})n\alpha
(\omega )}{\epsilon _{\infty }\sigma ^{3}+I_{1}(n\sigma ^{3})\alpha (\omega )
},
\end{equation}
where $I_{0}(x)$ and $I_{1}(x)$ are analytical functions depending
on dimensionless density $x=n\sigma ^{3}$ \cite{pratt}. Replacing Eq.(4) in
Eq.(3) leads to a quadratic equation, which allows
complete calculations of both the real and the imaginary part of
$\alpha(\omega)$. Our input phenomenological parameters are
1) $\sigma=2r_c=3.2$ \AA\ \cite{jortner1} which remains fixed for all
our calculations, 2) $\epsilon _{\infty }$=1.76 \cite{epsT}, and 3)
$\omega_0(T)$ extracted from experimental data. This last parameter
induces an implicit temperature dependence
of $\alpha(\omega)$. From \cite{omegaT}, we take $\omega_0(T=-70^{o}
\text{C})=0.9$ eV and $\partial \omega _{0}(T)/\partial T=-2.2\cdot
10^{-3}$ eV/K for higher temperature
($-70^{o}\text{C}<T<+70^{o}\text{C}$). The static dielectric
constant  in Fig. 3a is calculated by taking  $\epsilon
_{s}(T=-70^{o}$C$)$=25, and $\partial \epsilon _{s}/\partial T=-0.1\
$K$^{-1}$ \cite{epsT}.

\subsection*{Stability of delocalized states}

In a metallic state, the solution represents a plasma consisting of
a degenerate electron gas strongly coupled with the ions dissolved
in ammonia. Although a microscopic study of the system is still
beyond possibilities of current methods, low temperatures and low
metal concentrations of MAS simplify our analysis. We treat the
influence of the solvent simply as a screening effect of the
interactions, but we use different dielectric constants for the
interacting electron gas and the ionic potential, because the latter
are additionally screened by the orientational polarization of
solvent molecules. We characterize the plasma by dimensionless
parameters $r_{s}=(4\pi n/3)^{-1/3}a_{B}^{-1}$, $ \Gamma _{e}=\beta
e^2/\epsilon _{\infty} a_{B} r_{s}$, and $\Gamma _{i}=\beta e^2
/\epsilon _{s} a_{B} r_{s}$, where $a_{B}=\hbar ^{2}/me^{2}$. We can
express the change of the free energy caused by the
dissolution of metal atoms in terms of the dimensionless parameters,
and write the change $\Delta f$ per electron (or per metal atom) as
the sum of the electron, the ion, and the electron-ion contributions
\begin{equation}
\Delta f(n)=\Delta f_{e}(\Gamma _{e})+\Delta f_{i}(\Gamma _{i})+\Delta
f_{ie}(\Gamma _{e},\Gamma _{i},r_{s},a_{i}),
\end{equation}
where $a_{i}$ is a parameter related with the short-range
electron-ion pseudopotential, which takes into account deviations from Coulomb
interactions between electrons and ions. Because $\Gamma _{e}\gg
\Gamma _{i}$, it may be checked that the electron gas gives the main
contribution to $ \Delta f(n)$, whereas the ionic
contribution is only a correction. We can thus ignore thermal effects for simplified evaluations. We
take the expression obtained for the stabilized jellium model as in
\cite{folias}:
\begin{equation}
\Delta f(n)=\frac{3k_{F}^{2}}{10}-\frac{3k_{F}}{4\pi \epsilon _{\infty}}+e_{c}(n)+
\frac{C_{M}}{\epsilon _{s}r_{s}}+a_{i}n,  \label{fermi}
\end{equation}
where $k_{F}=(3\pi ^{2}n)^{1/3}a_{B}$ is the Fermi wave vector, and
$ C_{m}\approx -0.89774$ is the Madelung constant in atomic unit
(a.u.). The first, the second, and the third terms in Eq.(6) are the
kinetic, the exchange, and the correlation contributions to the
total energy of the homogeneous electron gas with a positive jellium
background. The last two terms are corrections which take into
account the atomic nature of the cations.
The only difference with the stabilized jellium model \cite{folias}, resides in the use of the dielectric
constant in the relevant contributions. The local density
approximation \cite{Pedrew} was applied to calculate $e_{c}(n)$. The main difficulty is to
evaluate $a_i$. In simple metals, the parameter $a_{i}$ may be
estimated as $\tilde{a}_{i}\approx 2\pi R^{2}/3$, where $R$ is the
ion-core radius related with the atomic number of the ion. Advanced
models treating smooth continuous pseudopotentials \cite{folias}
yield numerical corrections to this trend. We apply $\tilde{a}_{i}$
derived from \cite{folias} and take the screening effect of solvent
into account, i.e. $a_{i}=\tilde{ a}_{i}/\epsilon _{\infty}$.
Importantly, we here use $\epsilon_{\infty}$ because $a_i$
represents the short-range part of the interaction between electrons
and ions. As a result, we find $a_{i}=$10.8 and 26.1 a.u. for Na and
Cs respectively. Numerical deviations of about 10 percent from these
values do not change our  results significantly. With
the use of Eq.(6) we calculated the electronic part of the reduced isothermal compressibility $%
\kappa_{F} \sim [n\partial ^{2}( n\beta\Delta f)/\partial
n^{2}]^{-1}$ and found  the concentration $n_{c2}$ below which the
electron gas is unstable (Fig. 4).

\begin{acknowledgements}
G.N.Ch. thanks the Leverhulme Trust for partial support of this work. P.Q. thanks B.K. Chakraverty and D. Mayou for many discussions concerning
MAS. E-mail: pascal.quemerais@grenoble.cnrs.fr and genchuev@rambler.ru
\end{acknowledgements}

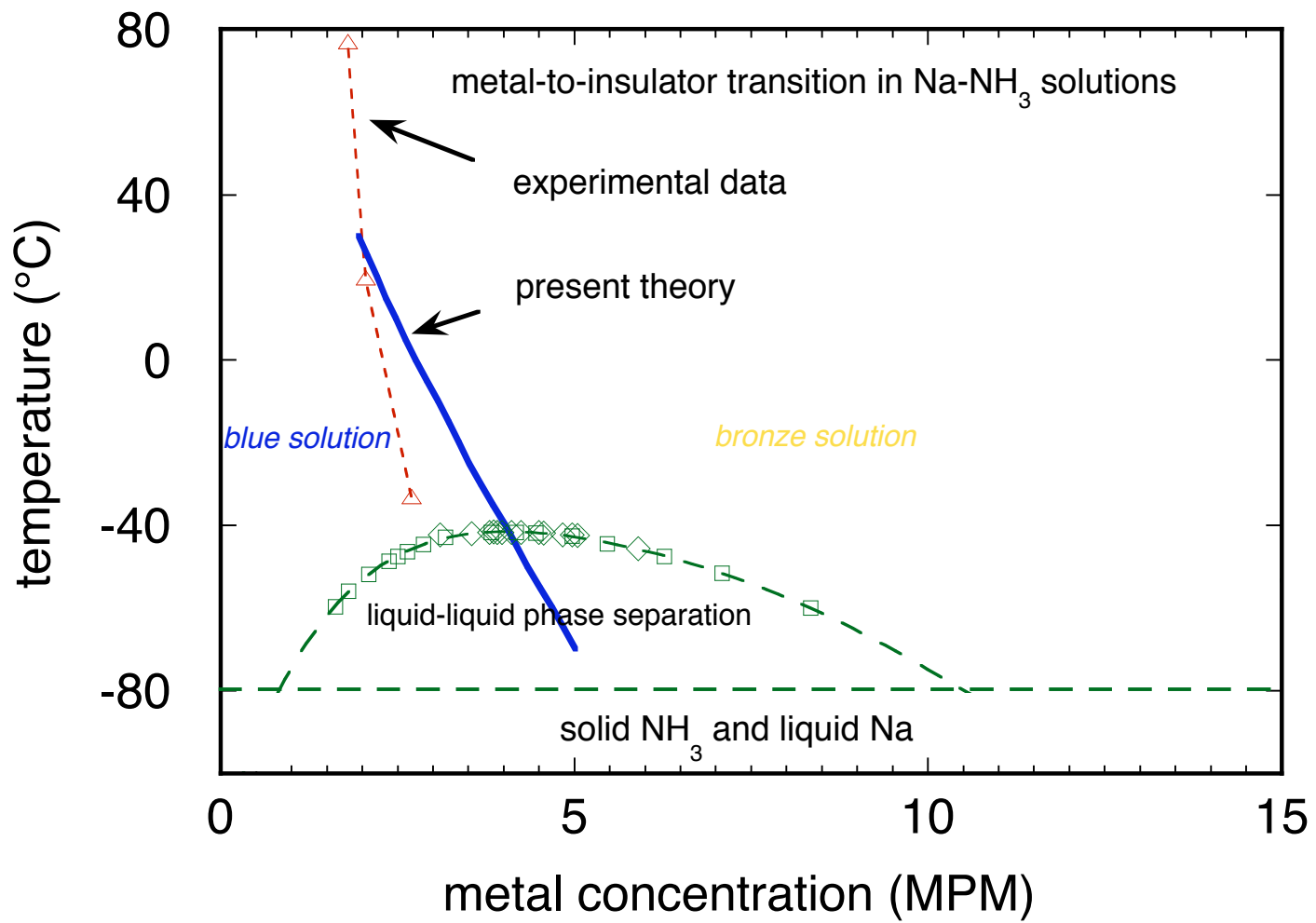
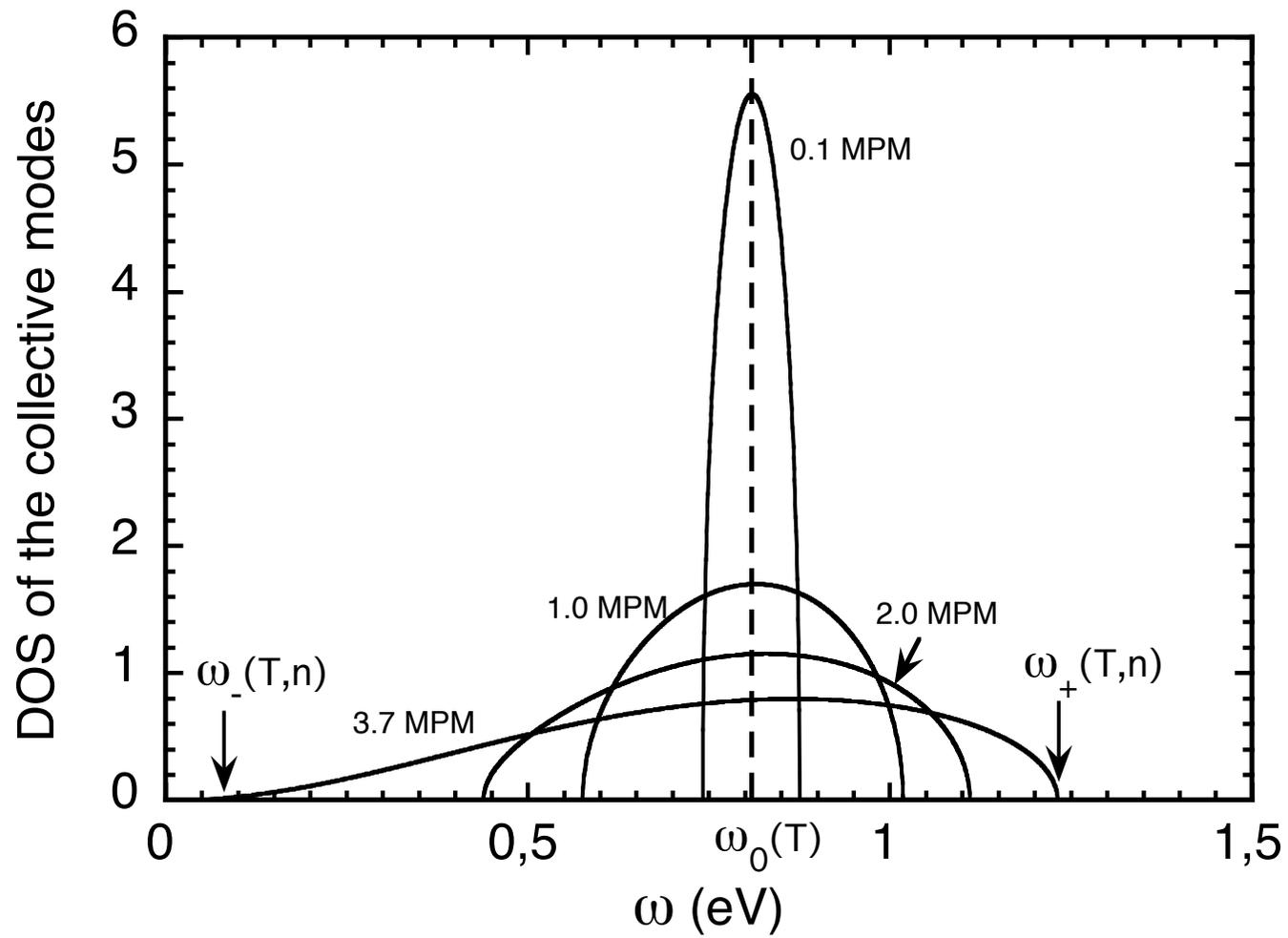
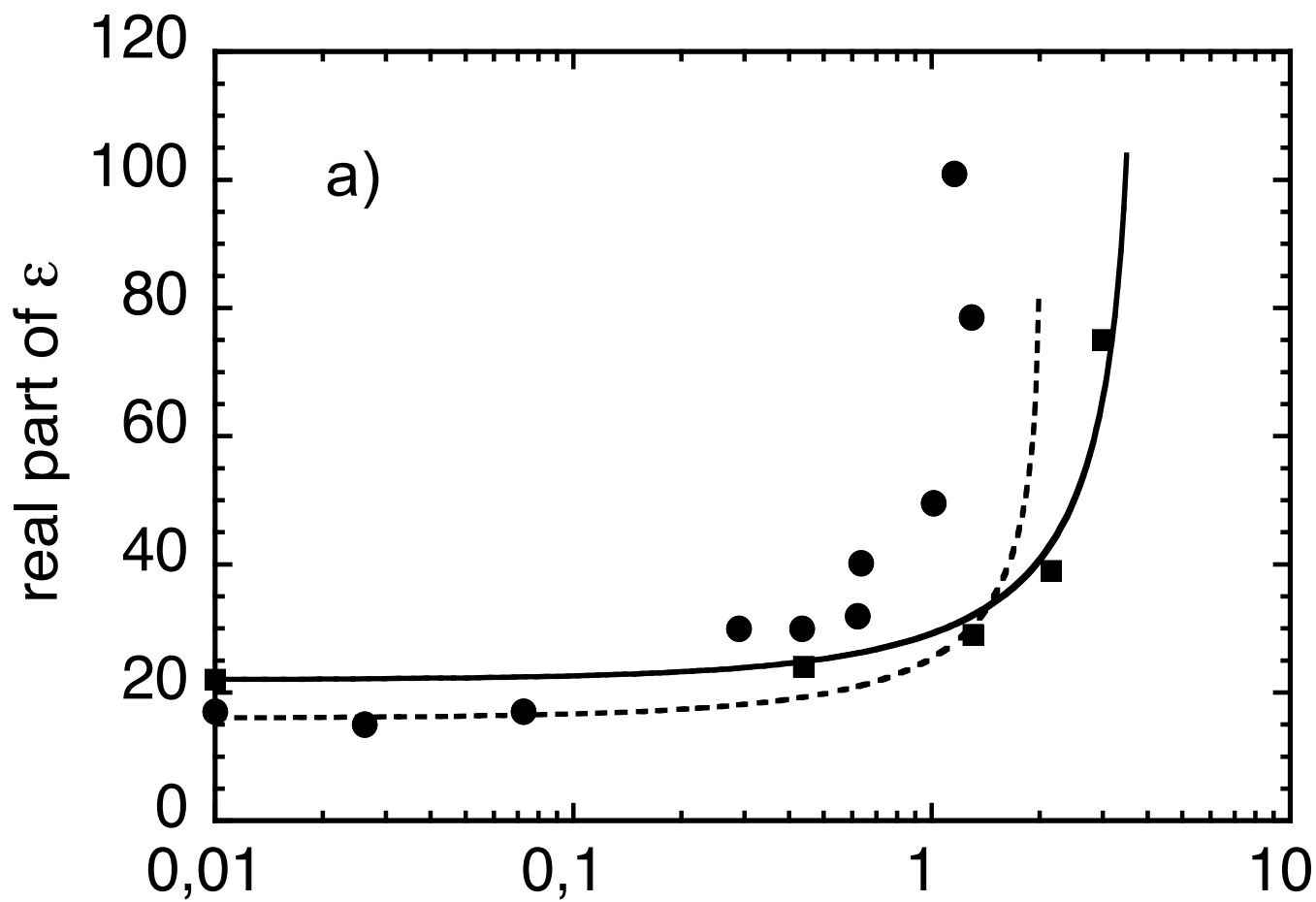
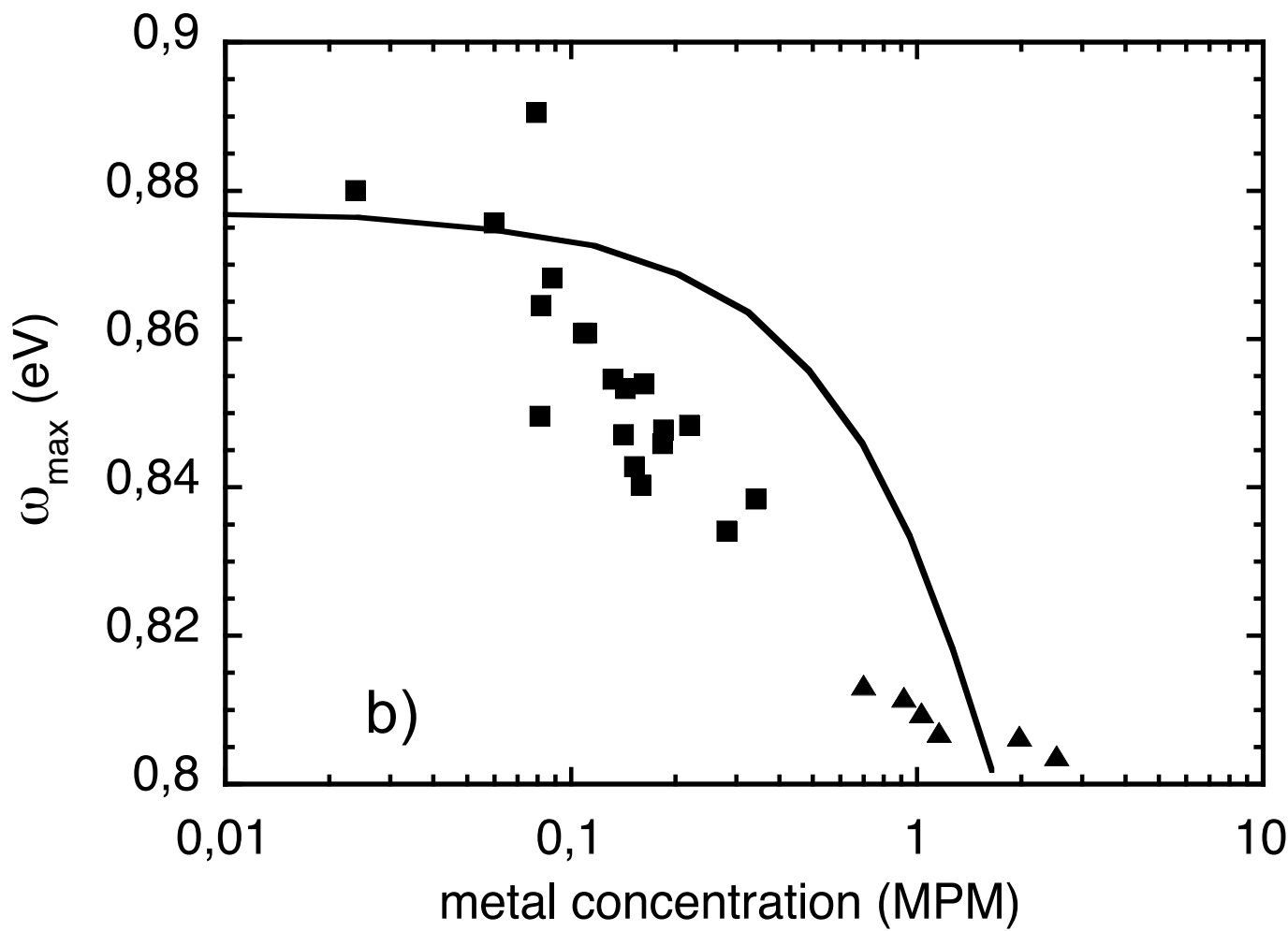
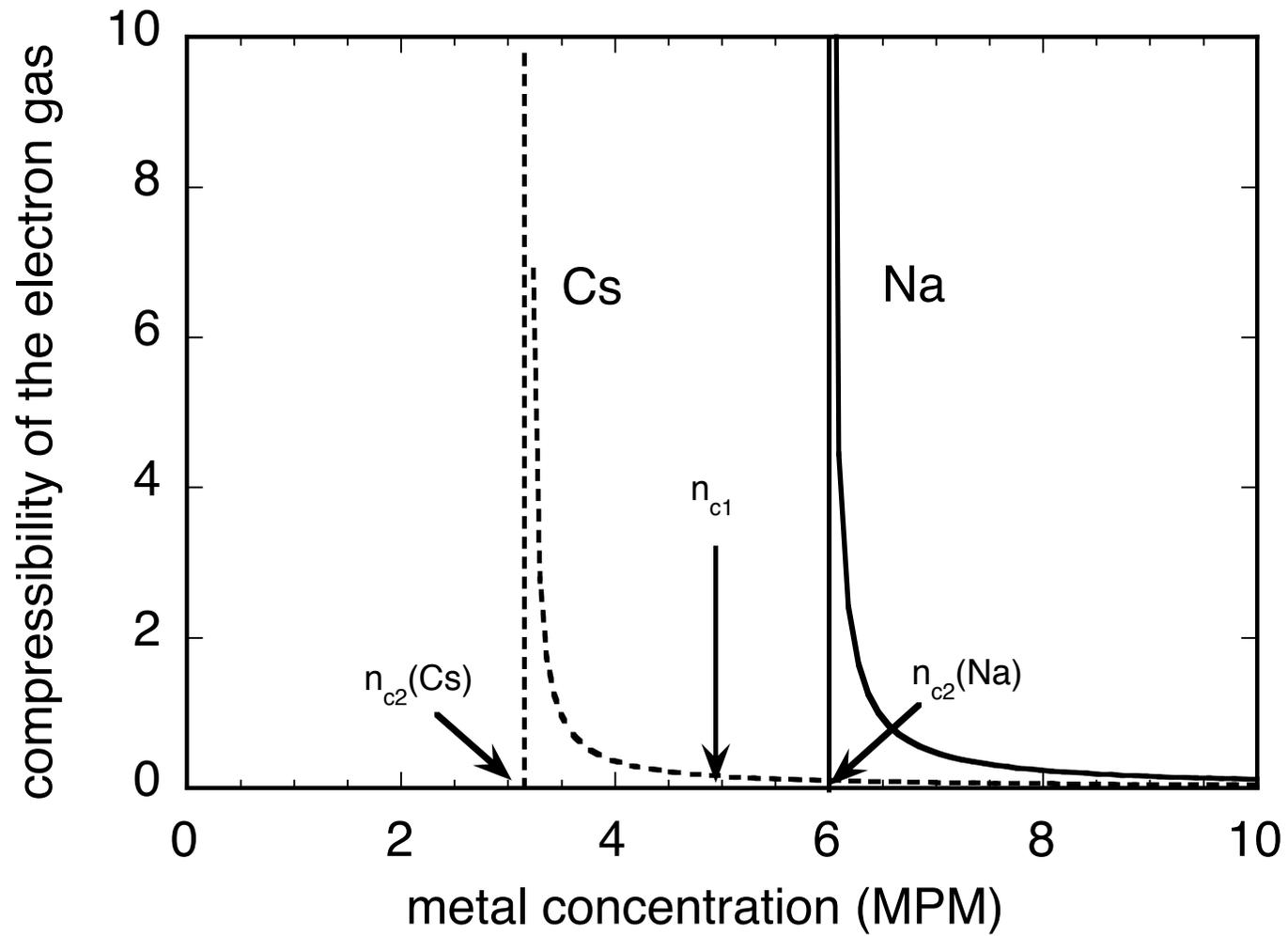
\end{document}